\documentclass[prd,onecolumn]{revtex4}

\usepackage{latexsym}
\usepackage{graphicx}
\usepackage{graphics}
\usepackage{epsfig}

\begin{document}

\title{Effects from inhomogeneities in the chiral transition}
\author{Bruno G. Taketani and Eduardo S. Fraga}
\affiliation{Instituto de F{\'\i}sica, Universidade 
Federal do Rio de Janeiro \\
C.P.~68528, Rio de Janeiro, RJ 21941-972, Brazil}

\begin{abstract}
We consider an approximation procedure to evaluate 
the finite-temperature one-loop fermionic density 
in the presence of a chiral background field which systematically 
incorporates effects from inhomogeneities in the chiral field through a 
derivative expansion. We apply the method to the case of a simple low-energy 
effective chiral model which is commonly used in the study of the 
chiral phase transition, the linear $\sigma$-model coupled to quarks. 
The modifications in the effective potential and their consequences for 
the bubble nucleation process are discussed.
\end{abstract}


\date{\today}
\maketitle



\section{Introduction}

It is commonly accepted that QCD at sufficiently high temperatures 
undergoes a phase transition to a new state of matter, the quark-gluon
plasma (QGP), which was presumably present in the early universe 
\cite{qgp,cosmo}. Compelling lattice QCD results corroborate this belief 
\cite{Karsch:2001vs}, and experiments in ultra-relativistic heavy-ion 
collisions \cite{rhic,QM2001} at BNL-RHIC have recently shown data that 
clearly point to a new state of matter \cite{jets-rhic}.

To model the mechanism of chiral symmetry breaking present in QCD, and to 
study the dynamics of phase conversion after a temperature-driven chiral 
transition, one can resort to low-energy effective models 
\cite{quarks-chiral,ove,Scavenius:1999zc,Scavenius:2000qd,
Scavenius:2001bb,Ignatius:1993qn,paech,Aguiar:2003pp,polyakov,explosive}. 
In particular, to study the mechanisms of bubble nucleation and 
spinodal decomposition in a hot expanding plasma \cite{Csernai:1992tj}, 
it is common to adopt the linear $\sigma$-model
coupled to quarks \cite{gellmann}, where the latter comprise the 
hydrodynamic degrees of freedom of the system. 
The gas of quarks provides a thermal bath in which the long-wavelength
modes of the chiral field evolve, and the latter plays the role of an order
parameter in a Landau-Ginzburg approach to the description of the chiral
phase transition \cite{Scavenius:2001bb,Ignatius:1993qn,paech,Aguiar:2003pp}. 
The standard procedure is then integrating over the fermionic degrees of 
freedom, using a classical approximation for the chiral field, to obtain a formal 
expression for the thermodynamic potential:
\begin{equation}
\Omega(T,\mu,\phi)=
V(\phi) - \frac{T}{{\cal V}}
\ln\det\left[\frac{(G_E^{-1}+M(\phi))}{T}\right]\; ,
\label{TDpot}
\end{equation}
where $V(\phi)$ is the classical self-interaction potential for the 
bosonic sector, $G_E$ is the fermionic Euclidean propagator, $M(\phi)$ 
is the effective fermion mass in the presence of the chiral field 
background, $T$ is the temperature and ${\cal V}$ is the 
volume of the system. From the thermodynamic potential (\ref{TDpot}), one 
can obtain all the physical quantities of interest. 

To actually compute correlation functions and thermodynamic 
quantities, one has to evaluate the fermionic determinant that results 
from the functional integration over the quark fields within some 
approximation scheme. Alternatively, one can consider the fermionic 
density which will appear as a source term in the equation of motion 
for the chiral field. In the case of one-dimensional systems 
one can often resort to exact analytical methods, such as the 
inverse scattering technique \cite{Fraga:1994xd}. 
In practice, however, the determinant is usually calculated to one-loop 
order assuming a homogeneous and static background 
field \cite{kapusta-book}. Nevertheless, for a system that is in the process 
of phase conversion after a chiral transition, one expects inhomogeneities 
in the chiral field configuration due to fluctuations to play a major 
role in driving the system to the true ground state. Hence, their effects 
should in principle be included in the computation of the fermionic determinant.

In the case of high-energy heavy ion collisions, hydrodynamical 
studies have shown that significant density inhomogeneities may 
develop dynamically when the chiral transition to the broken symmetry 
phase takes place \cite{paech} (see also \cite{Ignatius:1993qn} for 
an analysis in a different context). Their pattern and intensity might 
indeed provide some insight on the nature of the transition as well as on 
the location of an eventual critical point. If the freeze-out in heavy ion 
collisions occurs shortly after a first-order chiral transition, inhomogeneities 
generated during the late stages of the nonequilibrium evolution of the 
order parameter might leave imprints on the final spatial distributions 
and even on the integrated, inclusive abundances \cite{licinio}.

In this paper we consider an approximation procedure to evaluate 
the finite-temperature fermionic density in the presence 
of a chiral background field which systematically incorporates 
effects from inhomogeneities in the bosonic field through a 
gradient expansion. The method is valid for the case in which 
the chiral field varies smoothly, and allows one to extract information 
from its long-wavelength behavior, incorporating corrections order 
by order in the derivatives of the field \cite{fraser}. This approach 
has been successfully used to treat systems of low-dimensionality at 
zero temperature in condensed matter physics \cite{sakita}. Here 
we consider a three-dimensional system at finite temperature. 
We apply the method to the case of the linear $\sigma$-model 
coupled to quarks, which provides a convenient framework for the 
study of bubble nucleation and spinodal decomposition in the case 
of a first-order chiral transition. Nevertheless, the results presented 
below are quite general and may be of interest also in cosmology 
or in condensed matter systems.

The paper is organized as follows. Section II presents briefly the 
low-energy effective model adopted in this paper. In Section III 
we introduce the method to incorporate systematically 
effects from inhomogeneities in the chiral field in the computation of 
the fermionic density. Results for the (well-known) leading term 
and for the first non-trivial corrections are discussed in Section IV. 
There, we also consider the modifications undergone by the effective 
potential and their consequences to the process of nucleation. 
Section V contains our final remarks. 


\section{Effective model}

Let us consider a scalar field $\phi$ coupled to fermions $\psi$ 
according to the Lagrangian 
\begin{equation}
\mathcal{L} = \overline{\psi}[i\gamma^{\mu}\partial _{\mu} + \mu\gamma^0 -
M(\phi)]\psi + \frac{1}{2}\partial_{\mu}\phi ~\partial^{\mu}\phi - V(\phi)\; .
\label{lagrangian}
\end{equation}
where $\mu$ is the fermionic chemical potential, $M(\phi)$ is the effective 
mass of the fermions and $V(\phi)$ is a self-interaction potential for the 
bosonic field.

In the case of the linear $\sigma$-model coupled to quarks, $\phi$ represents 
the $\sigma$ direction of the chiral field $\Phi=(\sigma,\vec{\pi})$, where 
$\pi^i$ are pseudoscalar fields playing the role of the pions, which we drop 
here for simplicity. The pion directions play no major role in the 
process of phase conversion we have in mind, as was argued 
in Ref. \cite{Scavenius:2001bb}, so we focus on the sigma 
direction in what follows. However, the coupling of pions to 
the quark fields might be quantitatively important in the computation 
of the fermionic determinant inhomogeneity corrections. This issue 
makes the computation technically more involved and will be addressed 
in a future publication. 
The field $\psi$ plays the role of the constituent-quark field $q=(u,d)$, and 
$\mu=\mu_q$ is the quark chemical potential. The ``effective mass'' 
is given by $M(\phi)=g|\sigma|$, 
and $V(\Phi)=(\lambda^2/4)(\sigma^2+\vec{\pi}^2-v^2)^2-h_q\sigma$ is the 
self-interaction potential for $\Phi$. The parameters above are
chosen such that chiral $SU_{L}(2) \otimes SU_{R}(2)$ symmetry is
spontaneously broken in the vacuum. The vacuum expectation values of the
condensates are 
$\langle\sigma\rangle =\mathit{f}_{\pi}$ and $\langle\vec{\pi}\rangle =0$, 
where $\mathit{f}_{\pi}=93$~MeV is the pion decay constant.
The explicit symmetry breaking term is due to the finite current-quark
masses and is determined by the PCAC relation, giving 
$h_q=f_{\pi}m_{\pi}^{2}$, where $m_{\pi}=138$~MeV is the pion mass. This
yields $v^{2}=f^{2}_{\pi}-{m^{2}_{\pi}}/{\lambda ^{2}}$. The value of 
$\lambda^2 = 20$ leads to a $\sigma$-mass, 
$m^2_\sigma=2\lambda^{2}f^{2}_{\pi}+m^{2}_{\pi}$, equal to 600~MeV. 
In mean field theory, the purely bosonic part of this Lagrangian 
exhibits a second-order phase transition~\cite{Pisarski:1984ms} 
at $T_c=\sqrt{2}v$ if the explicit symmetry breaking term, $h_q$, 
is dropped. For $h_q\ne 0$, the transition becomes a smooth crossover 
from the restored to broken symmetry phases. For $g>0$, one has 
to include a finite-temperature one-loop contribution from the quark 
fermionic determinant to the effective potential as indicated in 
Eq. (\ref{TDpot}). When the coupling between quarks and the chiral field, 
$g$, is large enough, the system exhibits a first-order phase transition even 
at $\mu =0$ \cite{Scavenius:1999zc,Scavenius:2001bb,paech}. When we 
decrease $g$, the strength of this first-order transition is weakened. At 
$g\approx 3.7$, the latent heat vanishes and we have a second-order critical 
point at $\mu=0$. In what follows we keep the explicit symmetry breaking 
term $h_q\sigma$ and consider the case $g=5.5$, where the 
first-order line goes all the way down to $\mu=0$, since we are mainly 
concerned with the effects from inhomogeneities in the process of 
homogeneous nucleation.

The Euler-Lagrange equation for static chiral field configurations 
contains a term which represents the fermionic density, $\rho~$:
\begin{equation}
\nabla^2\phi=\frac{\partial V}{\partial\phi}+g\rho(T,\mu,\phi)\; ,
\label{euler-lagrange}
\end{equation}
and the density of fermions at a given point $\vec{x}_0$ has the form 
\begin{equation}
\rho(\vec{x}_0)=Sp \left\langle\vec{x}_0 
\left\vert \frac{1}{G_E^{-1}+M(\hat{x})}\right\vert 
\vec{x}_0\right\rangle \; ,
\label{density-Sp}
\end{equation}
where $\vert\vec{x}_0\rangle$ is a position eigenstate with eigenvalue 
$\vec{x}_0$, and $Sp$ represents a trace over fermionic degrees of 
freedom, such as color, spin and isospin. 

Assuming a homogeneous background field, one can compute
the one-loop fermionic density in a simple way \cite{kapusta-book}.
In this case, the correction coming from the integration over 
the fermions can be directly incorporated into an effective potential 
for the chiral field, as will be shown below. However, perfect 
homogeneity is a very strong hypothesis if one is interested in the 
dynamics of a phase transition. On the other hand, the correct 
determinant would have to be computed with an arbitrary profile 
for the background field. 
In a few examples, one can do it formally for one-dimensional systems
\cite{Fraga:1994xd}. For higher dimensions, however, one must
adopt  some approximation scheme to take into account
inhomogeneity effects. In the next section, we present a framework
to incorporate systematically derivative corrections to the
density $\rho(\vec{x})$. The only assumption made on the
behavior of  the background field is that it varies very smoothly.


\section{Inhomogeneity corrections}

In order to take into account inhomogeneity effects of the chiral background 
field, $\phi$, encoded in the position dependence of $M$ in (\ref{density-Sp}), 
we resort to a derivative expansion as explained below.

In momentum representation, the expression for the 
fermionic density assumes the form
\begin{equation}
\rho(\vec{x}_0)=Sp~ 
T\sum_n \int \frac{d^3k}{(2\pi)^3}
e^{-i\vec{k}\cdot\vec{x}_0} ~
\frac{1}{\gamma^0 (i\omega_n+\mu)-\vec{\gamma}\cdot\vec{k}+M(\hat{x})}
~e^{i\vec{k}\cdot\vec{x}_0} \; ,
\end{equation}
where $\omega_n = (2n+1)\pi T$ are Matsubara frequencies for 
fermions \cite{kapusta-book}. 
One can transfer the $\vec{x}_0$ dependence to $M(\hat{x})$ through a 
unitary transformation, obtaining
\begin{equation}
\rho(\vec{x}_0)=Sp~ 
T\sum_n \int \frac{d^3k}{(2\pi)^3}
\frac{1}{\gamma^0 (i\omega_n+\mu)-\vec{\gamma}\cdot\vec{k}
+M(\hat{x}+\vec{x}_0)} \; ,
\end{equation}
where one should notice that $\vec{x}_0$ is a c-number, not an operator.

Now we expand $M(\hat{x}+\vec{x}_0)$ around $\vec{x}_0$:
\begin{eqnarray}
M(\hat{x}+\vec{x}_0)&\equiv& 
M(\vec{x}_0)+\Delta M(\hat{x},\vec{x}_0)=\nonumber \\ 
&& M(\vec{x}_0)+\nabla_i M(\vec{x}_0)\hat{x}^i +
\frac{1}{2}\nabla_i \nabla_j M(\vec{x}_0)\hat{x}^i \hat{x}^j + 
\cdots \; ,
\end{eqnarray}
and use $\hat{x}^i=-i\nabla_{k_i}$ to write
\begin{equation}
\rho(\vec{x}_0)=Sp~ T\sum_n \int \frac{d^3k}{(2\pi)^3} 
\frac{1}{\gamma^0(i\omega_n+\mu)-\vec{\gamma}\cdot\vec{k}+M(\vec{x}_0)}
\left[ 1+ \Delta M(-i\nabla_{k_i},\vec{x}_0) 
\frac{1}{\gamma^0(i\omega_n+\mu)-\vec{\gamma}\cdot\vec{k}+M(\vec{x}_0)}
\right]^{-1}\, .
\label{brackets}
\end{equation}

To study the dynamics of phase conversion after a chiral 
transition, one can focus on the long-wavelength properties 
of the chiral field. From now on we assume that the 
static background, $M(\vec{x})$, varies smoothly and fermions 
transfer a small ammount of momentum to the chiral field, so 
that $\Delta M / M << 1$. Under 
this assumption, we can expand the expression inside brackets 
in Eq. (\ref{brackets}) in a power series:
\begin{equation}
\rho(\vec{x})=Sp~ T\sum_n \int \frac{d^3k}{(2\pi)^3} 
\frac{1}{\gamma^0(i\omega_n+\mu)-\vec{\gamma}\cdot\vec{k}+M(\vec{x})}
~\sum_\ell (-1)^\ell 
\left[ \Delta M(-i\nabla_{k_i},\vec{x}) 
\frac{1}{\gamma^0(i\omega_n+\mu)-\vec{\gamma}\cdot\vec{k}+M(\vec{x})}
\right]^{\ell}\, . 
\label{expansion}
\end{equation}
Eq. (\ref{expansion}), together with
\begin{equation}
\Delta M(-i\nabla_{k_i},\vec{x})=
\nabla_i M(\vec{x})\left(\frac{1}{i}\right)\nabla_{k_i}+
\frac{1}{2}\nabla_i \nabla_j M(\vec{x})\left(\frac{1}{i}\right)^2 
\nabla_{k_i}\nabla_{k_j} + \cdots \; ,
\end{equation}
provides a systematic procedure to incorporate corrections brought about 
by inhomogeneities in the chiral field to the quark density, so that 
one can calculate 
$\rho(\vec{x})=\rho_0(\vec{x})+\rho_1(\vec{x})+\rho_2(\vec{x})+\cdots$ 
order by order in powers of the derivative of the background, $M(\vec{x})$.

The new corrections will bring higher-order derivatives to the
equation of motion for the chiral field. In particular, as will be
seen below, the first non-trivial inhomogeneity contribution will
modify the Laplacian term in Eq. (\ref{euler-lagrange}), and
can be seen as a correction to the surface tension in the process
of bubble nucleation.

This is a quite general method to approximate the fermionic density 
and could be used in a variety of low-energy effective field theory models 
for the study of the dynamics of the chiral transition. In the next section we 
apply this method to the case of the linear $\sigma$-model coupled 
to quarks.


\section{Results}

\subsection{Leading term}

The leading-order term in this gradient expansion for $\rho(\vec{x})$ 
can be calculated in the standard fashion \cite{kapusta-book} and yields 
the well-known mean field result for the scalar quark density
\begin{equation}
\rho_0 =  \nu_q \int \frac{d^3k}{(2\pi)^3} 
\frac{M(\phi)/E_k(\phi)}{e^{[E_k(\phi)-\mu_q]/T}+1} + 
(\mu_q \to -\mu_q) \; ,  
\label{rho0}
\end{equation}
where $\nu_q=12$ is the color-spin-isospin degeneracy factor, 
$E_k(\phi)=(\vec{k}^2+M^2(\phi))^{1/2}$, and 
$M(\phi)=g|\phi|$ plays the role of an effective mass for the quarks. 
The net effect of this leading term is correcting the potential for the chiral 
field, so that we can rewrite Eq. (\ref{euler-lagrange}) as
\begin{equation}
\nabla^2\phi=\frac{\partial V_{eff}}{\partial\phi} \; ,
\label{euler-lagrange2}
\end{equation}
where $V_{eff}= V(\phi)+V_q(\phi)$ and
\begin{equation}
V_q\equiv -\nu_q T \int \frac{d^3k}{(2\pi)^3} 
\ln\left( e^{[E_k(\phi)-\mu_q]/T}+1 \right)+
(\mu_q \to -\mu_q) \; .
\end{equation}
The potentials $V_q$ and $V_{eff}$ for several values of the temperature 
(at $\mu_q=0$) are displayed in Fig. 1 and Fig. 2, respectively, 
assuming $g=5.5$. In this case, the chiral phase transition  is of first order  
even for a vanishing chemical potential, and $T_c\approx 124~$MeV. The 
barrier for nucleation disappears at $T_{sp}\approx 108~$MeV, 
where the system reaches the spinodal line \cite{Scavenius:2001bb}.
\vspace{0.5cm}

\begin{figure}[htb]
\centerline{\hbox{\epsfig{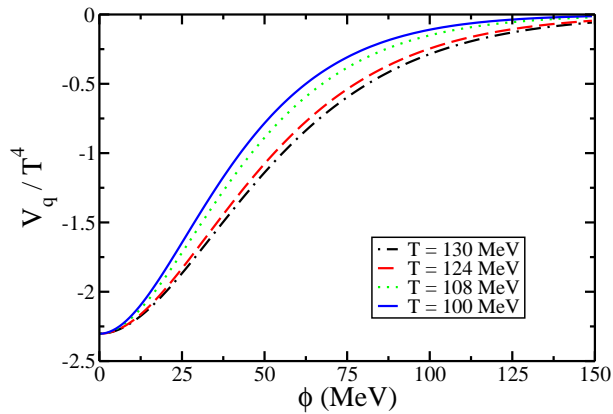}}}
\caption{$V_q(\phi)$ for different values of the temperature at 
$\mu_q=0$ and for $g=5.5$.}
\label{Veff}
\end{figure}

This kind of effective potential is commonly used as the coarse-grained 
thermodynamic potential in a phenomenological 
description of the chiral transition for an expanding 
quark-gluon plasma created in a high-energy heavy-ion collision 
\cite{Scavenius:1999zc,Scavenius:2000qd,Scavenius:2001bb,paech}. 
{}From the modified field equation (\ref{euler-lagrange2}), one can 
study, for instance, the phenomena of bubble nucleation and 
spinodal decomposition. However, the presence of a non-trivial 
background field configuration, {\it e.g.} a bubble, can in 
principle dramatically modify the Dirac spectrum \cite{rajaraman}, 
hence the determinant. In the case of condensed matter systems, where 
electronic doping often plays a major role, the presence of 
fermionic bound states can deeply affect the dynamics of the phase 
transition. This is the case in the presence of a bubble background, 
where besides unstable critical bubbles one can find metastable 
configurations, depending on the relative occupation of bound states, 
and a modification in the value of the nucleation 
rate \cite{Fraga:1994xd,yulu,polarons}. 
In the case of a chiral model, analogous effects can 
in principle appear for a nonzero quark chemical potential. 
In any case, one expects the effective potential to be 
modified by the effect of fluctuations of the chiral field on the fermionic 
density, which motivates the investigation of the next term 
in the expansion, which contains some information about the 
inhomogeneity of the bosonic field. 

\begin{figure}[htb]
\vspace{0.5cm}
\centerline{\hbox{\epsfig{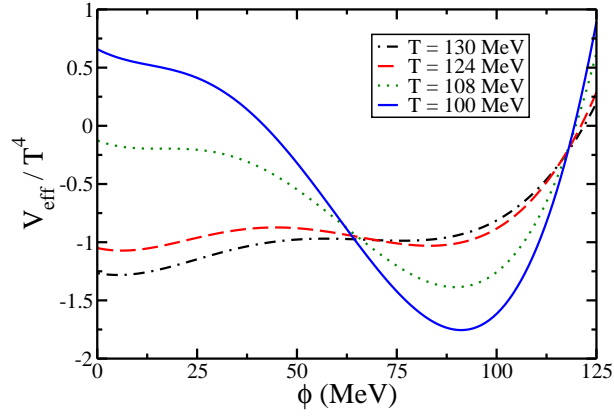}}}
\caption{$V_{eff}(\phi)$ in the $\sigma$ direction for different 
values of the temperature at $\mu_q=0$ and for $g=5.5$.}
\label{Vq-plot}
\end{figure}
%


\subsection{First corrections}

The next non-trivial term in the expansion contains two 
contributions: one coming from $\rho_1$ and another from $\rho_2$. 
This is due to the rearrangement of powers of the gradient 
operator. This term will correct the Laplacian piece in the chiral 
field equation. Dropping zero-temperature contributions which can 
be absorbed by a redefinition of the bare parameters in $V$, 
a long but straightforward calculation yields
\begin{equation}
(\rho_1 + \rho_2)=-(\nabla^2 M)~W_q(T,\mu_q,\phi)\; ,
\label{rho1+rho2}
\end{equation}
where 
\begin{equation}
W_q(T,\mu_q,\phi)=\frac{\nu_q}{2\pi^2}\int_0^\infty~
dk ~k^2 [H(E_k,T,\mu_q) 
+ H(E_k,T,-\mu_q)] \; ,
\end{equation}
\begin{eqnarray}
H(E_k,\mu_q,\beta)&=&-\;\beta^4\left(\frac{\hat{M}^4}{12}
\frac{\vec{k}^2}{E_k^5}\right)\left(e^{4\beta(E_k-\mu_q)}-11e^{3\beta(E_k-\mu_q)}
+11e^{2\beta(E_k-\mu_q)}-e^{\beta(E_k-\mu_q)}\right)n_F^5(E_k,\mu_q)\nonumber\\
&&\!\!\!\!-\beta^3\left(\frac{\hat{M}^2}{24}\frac{\vec{k}^2}{E_k^4}
+\frac{5\hat{M}^4}{6}\frac{\vec{k}^2}{E_k^6}\right)
\left(e^{3\beta(E_k-\mu_q)}-4e^{2\beta(E_k-\mu_q)}+e^{\beta(E_k-\mu)}\right)
n_F^4(E_k,\mu_q)\nonumber\\
&&\!\!\!\!-\beta^2\left(-\frac{\hat{M}^2}{16}\frac{1}{E_k^3}
-\frac{1}{24}\frac{\vec{k}^2}{E_k^3}+\frac{\hat{M}^2}{4}\frac{\vec{k}^2}{E_k^5}
+\frac{15\hat{M}^4}{4}\frac{\vec{k}^2}{E_k^7}\right)\left(e^{\beta(E_k-\mu_q)}-
e^{2\beta(E_k-\mu_q)}\right)n_F^3(E_k,\mu_q)\nonumber\\
&&\!\!\!\!-\beta\left(\frac{1}{4}\frac{1}{E_k^2}-\frac{3\hat{M}^2}{16}\frac{1}{E_k^4}
-\frac{1}{8}\frac{\vec{k}^2}{E_k^4}+\frac{5\hat{M}^2}{8}\frac{\vec{k}^2}{E_k^6}
+\frac{35\hat{M}^4}{4}\frac{\vec{k}^2}{E_k^8}\right)\; e^{\beta(E_k-\mu)}\;n_F^2(E_k,\mu_q)\nonumber\\
&&\!\!\!\!-\left(\frac{1}{8}\frac{1}{E_k^3}-\frac{3}{16}\frac{\hat{M}^2}{E_k^5}
-\frac{1}{8}\frac{\vec{k}^2}{E_k^5}+\frac{5\hat{M}^2}{8}
\frac{\vec{k}^2}{E_k^7}+\frac{35\hat{M}^4}{4}\frac{\vec{k}^2}{E_k^9}\right)n_F(E_k,\mu_q) \quad ,\nonumber\\
\end{eqnarray}
and $n_F(E_k,\mu_q)$ is the Fermi-Dirac distribution. 
The derivation of $H(E_k,T,\mu_q)$ is not particularly illuminating (a few 
steps are presented in the appendix). 
However, in the low-temperature limit, corresponding to $\beta M >> 1$, 
the integral above is strongly suppressed 
for high values of $k$, and the leading term has the much simpler form
\begin{equation}
W_q(T,\mu_q,\phi)\approx \frac{\nu_q}{\pi^2}\frac{\sqrt{2\pi}}{8}
e^{-\beta|M|} (\beta |M|)^{3/2} \, ,
\end{equation}
which gives a a better idea of the profile of the first inhomogeneity 
correction. One can already anticipate that it will be concentrated 
in the same region where the homogeneous correction was 
significant, i.e. $\beta M < 1$ (cf. Fig. 1), 
being exponentially suppressed for higher values of the field. In 
fact, a numerical study of the complete $W_q$ shows that this 
function is peaked around $\phi=0$ and non-negligible for 
$\beta|\phi|<1$ (see Fig. 3).
\begin{figure}[htb]
\centerline{\hbox{\epsfig{figure=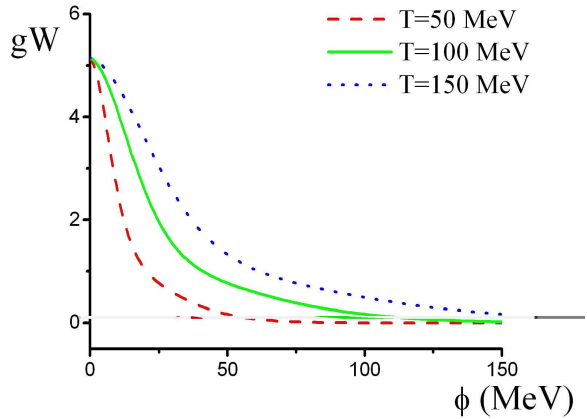,width=8cm}}}
\caption{$gW_q(\phi)$ for different 
values of the temperature at $\mu_q=0$ and for $g=5.5$.}
\label{gWq}
\end{figure}

The Euler-Lagrange equation for the chiral field up to this order in the gradient 
expansion reads
\begin{equation}
\nabla^2\phi=[1+gW_q(T,\mu_q,\phi)]^{-1}~
\frac{\partial V_{eff}(T,\mu_q,\phi)}{\partial\phi} \equiv 
V'_{in}(\phi)\; .
\end{equation}
Here we used the fact that $(1+gW_q)$ is a positive definite 
quantity, and {\it defined} a new ``effective potential'' that contains 
all the corrections up to this order in the gradient expansion, 
$V_{in}$. One should not confuse $V_{in}$ with the standard 
definition of the one-loop effective potential in field theory 
derived for a constant background \cite{itzykson}. Nevertheless, 
we keep the name effective potential for $V_{in}$ for convenience 
in the description of nucleation that follows.

The complete new effective potential can be obtained from our 
previous results by numerical integration. In order to proceed 
analytically, though, we choose to fit its derivative, which we 
know exactly up to this order, by a polynomial of the fifth degree. 
Actually, we know that the commonly used effective potential, 
$V_{eff}=V+V_q$, can hardly be distinguished from a fit with a 
polynomial of sixth degree in the region of interest for nucleation 
\cite{Fraga:2004hp}. Working with fits will 
be most convenient for using well-known results in the thin-wall 
approximation to estimate physical quantities that are relevant 
for nucleation, such as the surface tension and the free energy 
of the critical bubble. Results for the fits of $V'_{in}(\phi)$ for 
$\mu_q=0$ are shown in Fig. 4.
\begin{figure}[htb]
\vspace{0.5cm}
\centerline{\hbox{\epsfig{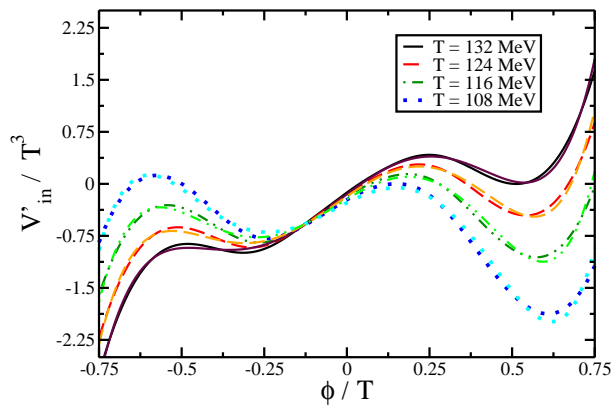}}}
\caption{Fits for $V'_{in}(\phi)$ for different 
values of the temperature at $\mu_q=0$ and for $g=5.5$. 
The lines shown in the box correspond to the exact results.}
\label{fits-derVin}
\end{figure}

We can now integrate analytically the polynomial approximation 
to the derivative of the complete effective potential. In Fig. 5 we 
display the curves for $V_{eff}$ and $V_{in}$ for a few values 
of temperature and $\mu_q=0$.

\begin{figure}[hbtp]
  \vspace{0.5cm}
  \centerline{\hbox{ \hspace{-0.2in}
    \includegraphics[angle=0,width=3in]{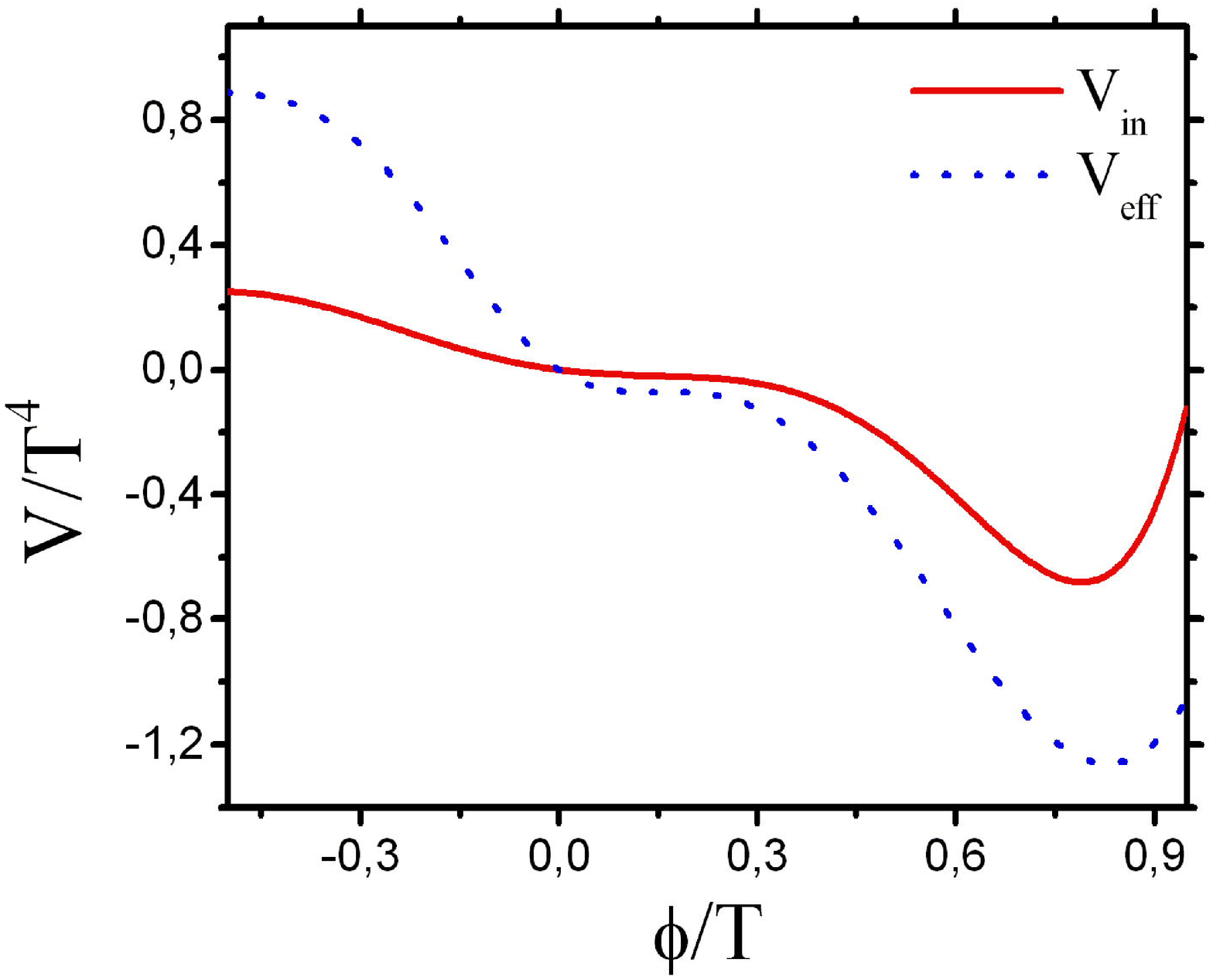}
    \hspace{0.25in}
    \includegraphics[angle=0,width=3in]{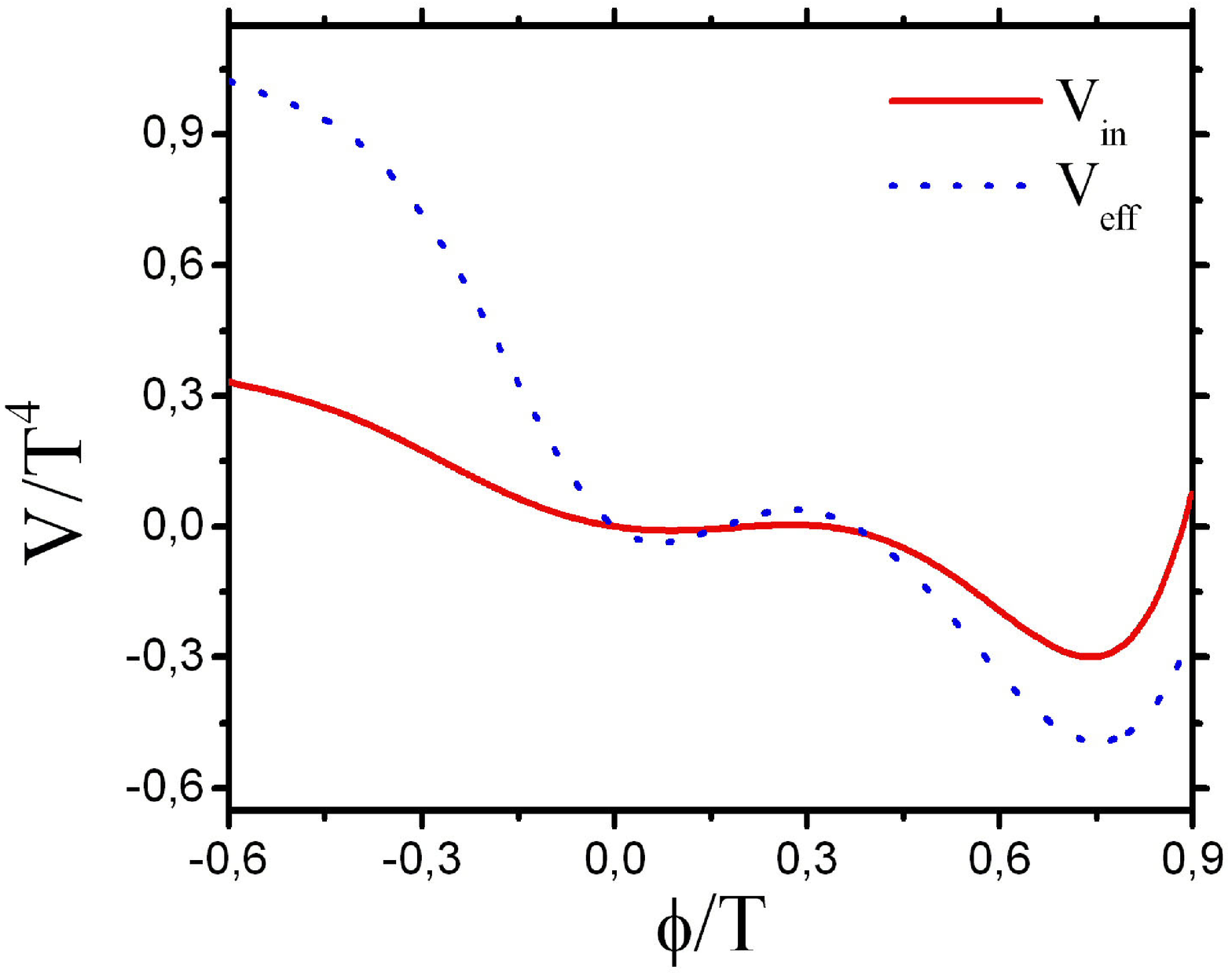}
    }
  }
  \vspace{-10pt}
  \hbox{\hspace{1.4in} (a) \hspace{2.97in} (b) }
  \vspace{9pt}

  \centerline{\hbox{ \hspace{-0.2in}
    \centerline{\includegraphics[angle=0,width=3in]{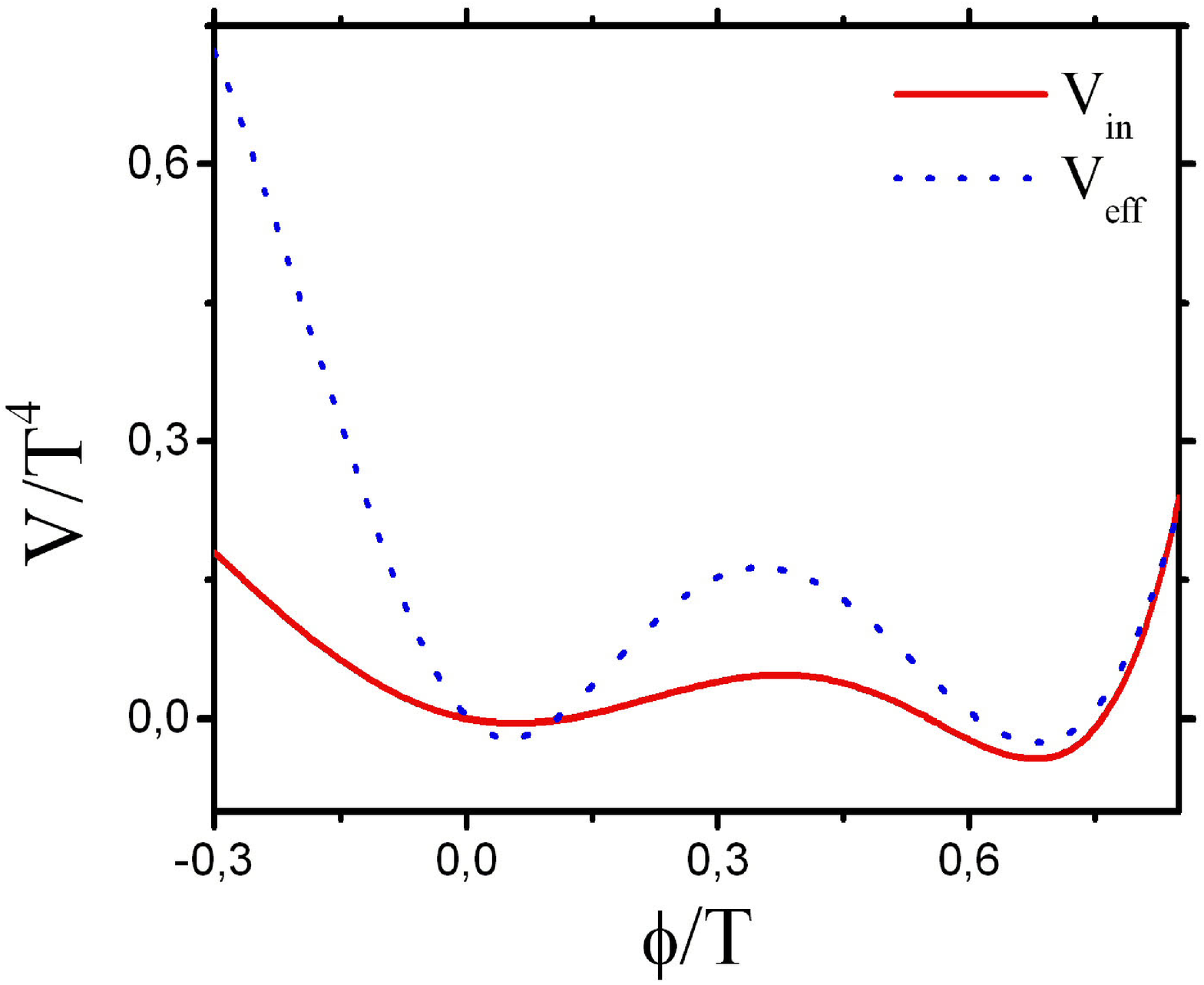}}
    \hspace{-3.8in}
    \centerline{\includegraphics[angle=0,width=3in]{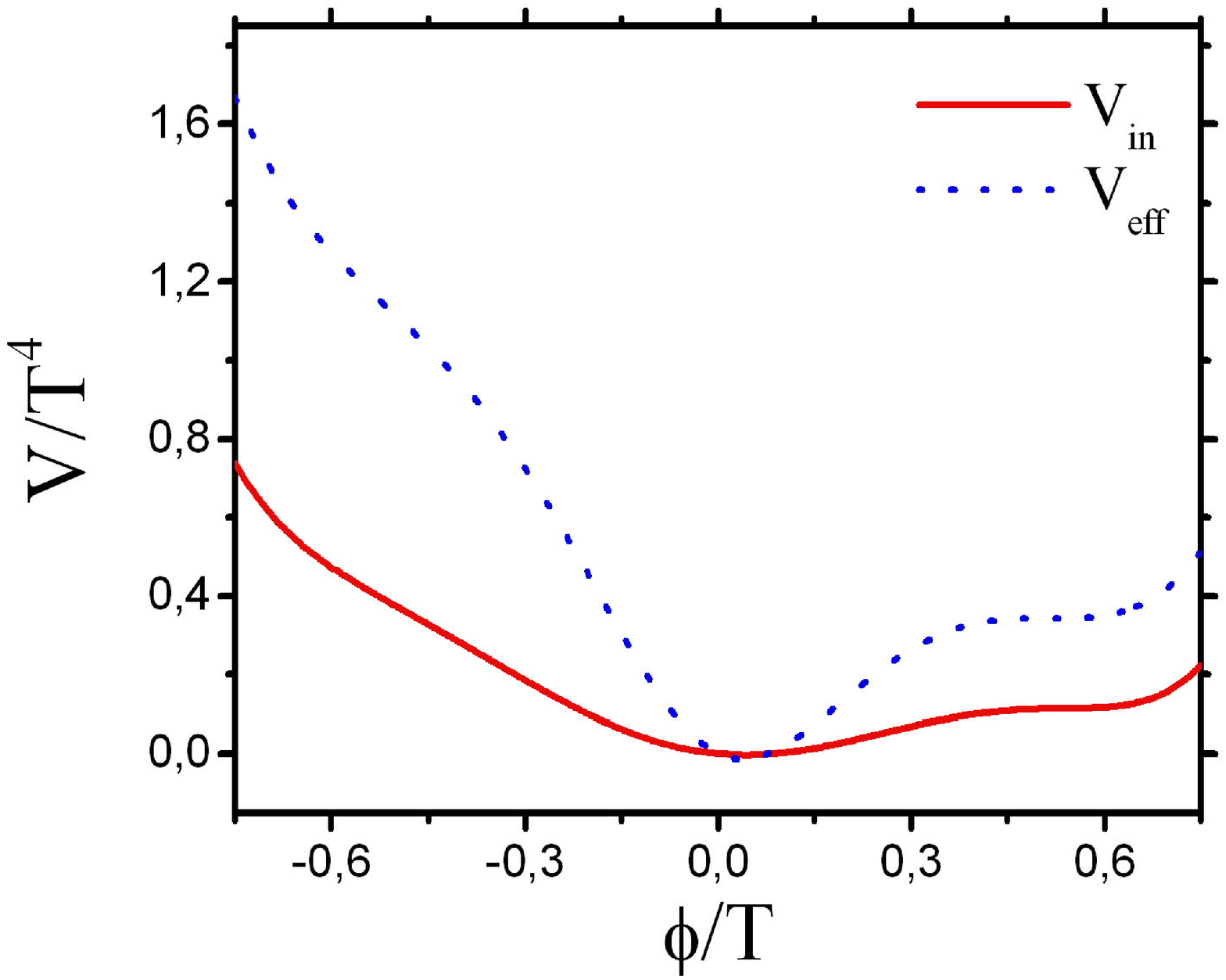}}
    }
  }
  \vspace{-160pt}
  \vspace{135pt}

  \vspace{15pt}
  \hbox{\hspace{1.4in} (c) \hspace{2.97in} (d) }
  \vspace{9pt}
  \label{g=3.3}

  \caption{$V_{eff}$ and $V_{in}(\phi)$ for different 
values of the temperature $T=[108~ {\rm (a)}, 116~ {\rm (b)}, 
124~ {\rm (c)}, 132~ {\rm (d)}]$ MeV 
at $\mu_q=0$ and for $g=5.5$.}
\label{Veff-Vin}
\end{figure}

{}From Fig. 5 one can notice a few consequences of the inhomogeneity 
correction. The first general effect is the smoothening of the effective 
potential. In particular, and most importantly, the barrier between 
the symmetric phase and the broken phase is significantly diminished, 
as well as the depth of the broken phase minimum, although we still 
have a first-order phase transition barrier. Therefore, one can 
expect an augmentation in the bubble nucleation rate. In principle, one 
should have better results from calculations within the thin-wall 
approximation. Also, the critical temperature moves up slightly. 


\subsection{Effects on nucleation}

Let us now consider the effects of the first inhomogeneity correction on 
the process of phase conversion driven by the nucleation of bubbles 
\cite{reviews}. To work with approximate analytic formulas, we follow 
Ref. \cite{Scavenius:2001bb} and express $V_{in}$ over the range 
$0 \le \phi \le T$ in the familiar Landau-Ginzburg form
\begin{equation}
V_{\rm eff} \approx \sum_{n=0}^4 a_n \, \Phi^n\quad.
\label{LandGinz}
\end{equation}
Although this approximation is obviously incapable of reproducing all three 
minima of $V_{in}$, this polynomial form is found to provide a good quantitative 
description of $V_{in}$ in the region of interest for nucleation, i.e. where 
the minima for the symmetric and broken phases, as well as  the barrier 
between them, are located. 

A quartic potential such as Eq.\,(\ref{LandGinz}) can always be rewritten in the form
\begin{equation}
U(\varphi)=\alpha ~ (\varphi^2-a^2)^2+j\varphi\quad.
\label{www}
\end{equation}
The coefficients above are defined as follows:
\begin{eqnarray}
\alpha &=& a_4\quad, \\
a^2 &=& \frac{1}{2}\left[ -\frac{a_2}{a_4}
+\frac{3}{8}\left(\frac{a_3}{a_4}\right)^2  \right]\quad, \\
j &=& a_4\left[ \frac{a_1}{a_4}-
\frac{1}{2}\frac{a_2}{a_4}\frac{a_3}{a_4}+
\frac{1}{8}\left(\frac{a_3}{a_4}\right)^3   \right]\quad, \\
\varphi&=&\phi + \frac{1}{4}\frac{a_3}{a_4} \ . 
\end{eqnarray}
The new potential $U(\varphi)$ reproduces the original $V_{in} ( \phi )$ 
up to a shift in the zero of energy.  We are interested in the effective 
potential only between $T_c$ and $T_{sp}$.  At $T_c$, we will have two 
distinct minima of equal depth.  This clearly corresponds to the choice $j = 0$
in Eq.\,(\ref{www}) so that $U$ has minima at $\varphi = \pm a$ and a maximum 
at $\varphi = 0$.  The minimum at $\varphi = -a$ and the maximum move closer 
together as the temperature is lowered and merge at $T_{sp}$.  Thus, 
the spinodal requires $j/\alpha a^3 = -8/3\sqrt{3}$ in Eq.\,(\ref{www}). 
The parameter $j/\alpha a^3$ falls roughly 
linearly from $0$, at $T=T_c$, to $-8/3\sqrt{3}$ at the spinodal.

The explicit form of the critical bubble in the thin-wall limit is then given 
by~\cite{Fraga:1994xd}
\begin{equation}
\varphi_c (r;\xi,R_c)=\varphi_f + \frac{1}{\xi\sqrt{2\alpha}}
\left[ 1-\tanh \left( \frac{r-R_c}{\xi} \right) \right]\quad,
\label{bubproftw}
\end{equation}
where $\varphi_f$ is the new false vacuum, $R_c$ is the radius of the 
critical bubble, and $\xi=2/m$, with $m^2\equiv U''(\varphi_f)$, is a measure 
of the wall thickness.  The thin-wall limit corresponds to $\xi/R_c\ll 1$ 
\cite{Fraga:1994xd}, which can be rewritten as $(3|j|/8\alpha a^3)\ll 1$.  
This small parameter has the value of $1/\sqrt{3}$ at the spinodal, which 
suggests that the thin-wall approximation might be qualitatively reliable for 
our purposes.  Nevertheless, it was shown in \cite{Scavenius:2001bb} 
that the thin-wall limit becomes very imprecise as one approaches the 
spinodal. In this vein, the analysis presented below is to be regarded as 
semi-quantitative. To be consistent we compare results from the homogeneous 
calculation to those including the inhomogeneity correction within the 
same approximation. 

In terms of the parameters $\alpha$, $a$, and $j$ defined above, we find 
\begin{eqnarray}
\varphi_{t,f} &\approx& \pm a - \frac{j}{8\alpha a^2}  \quad, \\
\xi &=& \left[ \frac{1}{\alpha (3\varphi_f^2-a^2)} \right]^{1/2}
\label{twcorlength}
\end{eqnarray}
in the thin-wall limit.  Determination of the critical radius requires 
the surface tension, $\Sigma$, defined as 
\begin{equation}
\Sigma\equiv \int_0^{\infty}{\rm d}r~\left( \frac{{\rm d}
\varphi_b}{{\rm d}r} \right)^2 
\approx \frac{2}{3\alpha\xi^3} \ .
\end{equation}
The critical radius then becomes $R_c = (2\Sigma/\Delta U)$, where
$\Delta U \equiv V(\phi_f)-V(\phi_t) \approx 2 a | j |$. 
The free energy of a critical bubble is finally given by
$F_b=(4\pi\Sigma/3)R_c^2$.
{}From knowledge of $F_b$, one can evaluate the nucleation rate 
$\Gamma \sim e^{-F_b/T}$. In calculating thin-wall properties, we shall 
use the approximate forms for $\varphi_t$, $\varphi_f$, $\Sigma$, 
and $\Delta U$ for all values of the potential parameters.

To illustrate the effect from the inhomogeneity correction, we compute 
the critical radius and $F_b/T$ for three different values of the temperature. 
For $T=108~$MeV, corresponding to the spinodal temperature, which is 
not modified by the first inhomogeneity correction, the corrected 
values are $R_c\approx 0.98~$fm and $F_b/T\approx 0.20$, as compared to 
$R_c\approx 1.1~$fm and $F_b/T\approx 0.9$ in the homogeneous case. 
The same computation for $T=116~$MeV yields 
$R_c\approx 2.15~$fm and $F_b/T\approx 1.14$, as compared to 
$R_c\approx 2.2~$fm and $F_b/T\approx 2.1$. At $T=124~$MeV, which 
corresponds to the critical temperature for the homogeneous case, 
the critical radius and $F_b/T$ diverge in the homogeneous computation, 
whereas $R_c\approx 35~$fm and $F_b/T\approx 394$ including 
inhomogeneities. The numbers above clearly indicate that the formation 
of critical bubbles is much less suppressed in the scenario with 
inhomogeneities, which will in principle accelerate the phase conversion 
process after the chiral transition.


\section{Summary and outlook}

We have introduced a systematic procedure to evaluate 
inhomogeneity corrections to the finite-temperature fermionic 
density in the presence of a chiral background field, which 
incorporates effects from fluctuations in the bosonic field through 
a gradient expansion at finite temperature and density. Higher-order 
contributions give more non-local corrections to the effective 
Euler-Lagrange equation for the chiral field, and the condition for 
the validity of the method is a smooth variation of the chiral field, 
which should be enough in the analysis of its long-wavelength 
behavior in the phase transition.

Incorporating the first inhomogeneity correction in the computation 
of the effective potential of the linear $\sigma$-model coupled to 
quarks, we found that the latter is significantly modified. Besides 
a general smoothening of the potential, the critical temperature 
moves upward and the hight of the barrier separating the symmetric 
and the broken phase vacua diminishes appreciably. As a direct 
consequence, the radius of the critical bubble goes down, as well 
as its free energy, and the process of nucleation is facilitated. 
Although the numbers presented above should be regarded as simple 
estimates, since they rely on a number of approximations, the 
qualitative behavior is clear. In a detailed quantitative 
analysis, one should not only integrate numerically the effective 
potential and relax the thin-wall approximation, but also 
include the pion-quark interaction in the computation of the 
fermionic density. We believe that the contribution from the 
pion sector will enhance the effect from inhomogeneities.

In all the discussion above, we intentionally ignored corrections
coming from bosonic fluctuations, which would result in a bosonic
determinant correction to the effective potential \cite{bosonic}.
To focus on the effect of an inhomogeneous background field on the
fermionic density, we treated the scalar field essentially as a  
``heavy'' (classical) field, whereas fermions were assumed to 
be ``light''.

Experimental signatures of inhomogeneities for high-energy 
heavy ion collisions were discussed, for instance, in 
Refs. \cite{paech,licinio}. In particular, inhomogeneities 
seem to favor an ``explosive'' scenario \cite{explosive}
for the phase conversion even at early stages of nucleation. 
However, one should first incorporate dissipation and noise 
effects, which tend to retard the explosion 
\cite{Fraga:2004hp}, before estimating the 
time scales involved. This analysis will be left for a 
future publication.


\begin{acknowledgments}
The authors are grateful to A. Dumitru for a critical reading of 
the manuscript and several suggestions. We also thank D.G. Barci, 
H. Boschi-Filho, C.A.A. de Carvalho and T. Kodama for discussions. 
This work was partially supported by CAPES, CNPq, FAPERJ and FUJB/UFRJ. 
\end{acknowledgments}


\appendix

\section{}

In this appendix we sketch the main steps to build the function $W_q(T,\mu_q,\phi)$ that 
corrects the Laplacian in the Euler-Lagrange equation for the chiral field.

The first inhomogeneity correction has contributions from $\rho_1$ and $\rho_2$.  The 
contribution coming from $\rho_1$ is proportional to $\nabla_i\nabla_j\phi$:
\begin{equation}
\rho_1^{(2)}(\vec{x})=-Tr_\gamma\;\int_K
\frac{\gamma^0k_0-\vec{\gamma}\cdot\vec{k}-
M(\vec{x})}{k_0^2-E_k^2}\left(\frac{1}{2}\nabla_i\nabla_j M(\vec{x}_0)\hat{x}^i\hat{x}^j\right)
\left(\frac{\gamma^0k_0-\vec{\gamma}\cdot\vec{k}-M(\vec{x})}{k_0^2-E_k^2}\right) \quad ,
\end{equation}
where we use a compact notation for the sum-integrals 
\begin{equation}
\int_K \equiv T\sum_n \int \frac{d^3k}{(2\pi)^3} \quad ,
\end{equation}
$K\equiv (k_0,\vec{k})=(i\omega_n-\mu_q,\vec{k})$, 
and $Tr_\gamma$ is a trace over Dirac gamma matrices. 

There is also a contribution proportional to $\nabla_i\phi\nabla_j\phi$ coming from $\rho_2$:
\begin{equation}
\rho_2^{(1)}(\vec{x})=Tr_\gamma\;
\int_K\frac{\gamma^0k_0-\vec{\gamma}\cdot\vec{k}-
M(\vec{x})}{k_0^2-E_k^2}\left[\nabla_i M(\vec{x})\hat{x}^i
\left(\frac{\gamma^0k_0-\vec{\gamma}\cdot\vec{k}-
M(\vec{x})}{k_0^2-E_k^2}\right)\right]^2 \quad .
\end{equation}

Up to this order in derivatives, we can write
\begin{eqnarray}
\rho_1+\rho_2&=&g\int\frac{d^3k}{(2\pi)^3}\nu_q 
\left\{\nabla^2\hat{M}(\vec{x})\left[F(E_k,\mu_q,\beta)+
F(E_k,-\mu_q,\beta)\right]+(\vec{k}\cdot\nabla)^2 M(\vec{x})
\left[G(E_k,\mu_q,\beta)+G(E_k,-\mu_q,\beta)\right]\right\}\quad ,
\nonumber \\
\end{eqnarray}
where $\beta = 1/T$, 
\begin{eqnarray}
F(E_k,\mu_q,\beta)\!&=&\!-\beta^2\left(-\frac{1}{2^4}\frac{M^2}{E_k^3}\right)
\left(\frac{e^{2\beta(E_k-\mu_q)}-e^{\beta(E_k-\mu_q)}}{(e^{\beta(E_k-\mu_q)}+1)^3}\right)
\nonumber\\
&&
-\beta\left(\frac{1}{2^2E_k^2}-\frac{3}{2^4}\frac{M^2}{E_K^4}\right)
\left(\frac{e^{\beta(E_k-\mu_q)}}{(e^{\beta(E_k-\mu_q)}+1)^2}\right)
\nonumber\\
&&
-\left(\frac{1}{2^3E_k^3}-\frac{3}{2^4}\frac{M^2}{E_K^5}\right)
\left(\frac{1}{(e^{\beta(E_k-\mu_q)}+1)}\right)
\end{eqnarray}
and
\begin{eqnarray}
G(E_k,\mu_q,\beta)&=&-\;\beta^4\left(\frac{1}{2^2}\frac{M^4}{E_k^5}\right)
\left(\frac{e^{4\beta(E_k-\mu_q)}-11e^{3\beta(E_k-\mu_q)}+
11e^{2\beta(E_k-\mu_q)}-e^{\beta(E_k-\mu_q)}}{(e^{\beta(E_k-\mu_q)}+1)^5}\right)
\nonumber\\
&&-\beta^3\left(\frac{1}{2^3}\frac{M^2}{E_K^4}+\frac{5}{2}\frac{M^4}{E_K^6}\right)
\left(\frac{e^{3\beta(E_k-\mu_q)}-4e^{2\beta(E_k-\mu_q)}+
e^{\beta(E_k-\mu_q)}}{(e^{\beta(E_k-\mu_q)}+1)^4}\right)
\nonumber\\
&&-\beta^2\left(-\frac{1}{2^3E_K^3}+\frac{3}{2^2}\frac{M^2}{E_K^5}+
\frac{45}{2^2}\frac{M^4}{E_K^7}\right)
\left(\frac{e^{2\beta(E_k-\mu_q)}-e^{\beta(E_k-\mu_q)}}{(e^{\beta(E_k-\mu_q)}+1)^3}\right)
\nonumber\\
&&-\beta\frac{3}{2^2}\left(-\frac{1}{2E_K^4}+\frac{5}{2}\frac{M^2}{E_K^6}+
35\frac{M^4}{E_K^8}\right)
\left(\frac{e^{\beta(E_k-\mu_q)}}{(e^{\beta(E_k-\mu_q)}+1)^2}\right)
\nonumber\\
&&-\frac{3}{2^2}\left(-\frac{1}{2E_K^5}+\frac{5}{2}\frac{M^2}{E_K^7}+35\frac{M^4}{E_K^9}\right)
\left(\frac{1}{(e^{\beta(E_k-\mu_q)}+1)}\right) \quad .
\end{eqnarray}

Rewriting the term proportional to $\vec{k}\cdot\nabla$ in a more convenient form, using 
$E_k^2=\vec{k}^2+M^2$, and exploring some symmetries in the integrands, it is 
straightforward to arrive at the final form:
\begin{equation}
\rho_1+\rho_2=\nabla^2\phi(\vec{x})\frac{g\nu_q}{2\pi^2}
\int dk\,\vec{k}^2\left[H(E_k,\mu_q,\beta)+H(E_k,-\mu_q,\beta)\right]\,\,\,,
\end{equation}
where
\begin{equation}
H(E_k,\mu_q,\beta)\equiv F(E_k,\mu_q,\beta)+\frac{1}{3}\vec{k}^2\,G(E_k,\mu_q,\beta)\,\,\, ,
\end{equation}
which gives the correction to the Laplacian.




\begin{thebibliography}{99}

\bibitem{qgp} 
E.V.~Shuryak, 
Phys.\ Rept.\ {\bf 61}, 71 (1980); 
K.~Kajantie and L.~McLerran, 
Ann.\ Rev.\ Nucl.\ Part.\ Sci.\  {\bf 37}, 293 (1987); 
B.~M\"uller, Rept.\ Prog.\ Phys.\ {\bf 58}, 611 (1995);
J.~Harris and B.~M\"uller, 
Ann.~Rev.~Nucl.~Part.~Sci.~{\bf 46}, 71 (1996); 
S.A.~Bass, M.~Gyulassy, H.~St\"ocker and W.~Greiner, 
J.~Phys.~{\bf G25}, R1 (1999).

\bibitem{cosmo} 
E.W. Kolb and M.S. Turner, 
\textit{The Early Universe} (Addison-Wesley, Redwood City, 1990).

\bibitem{Karsch:2001vs} 
F.~Karsch,
Nucl.\ Phys.\ A {\bf 698}, 199 (2002); 
E.~Laermann and O.~Philipsen,
Ann.\ Rev.\ Nucl.\ Part.\ Sci.\  {\bf 53}, 163 (2003).

\bibitem{rhic} 
J.~Harris and B.~M\"uller, 
Ann.~Rev.~Nucl.~Part.~Sci.~\textbf{46}, 71 (1996).

\bibitem{QM2001} 
Proc. of Quark Matter 2004, 
J. Phys. G \textbf{30}, S633-S1425 (2004).

\bibitem{jets-rhic} 
J.~Adams {\it et al.}  [STAR Collaboration],
Phys.\ Rev.\ Lett.\  {\bf 91}, 072304 (2003).

\bibitem{quarks-chiral} 
L.~P.~Csernai and I.~N.~Mishustin, Phys.\ Rev.\
Lett.\ \textbf{74}, 5005 (1995); 
A.~Abada and J.~Aichelin, 
Phys.\ Rev.\ Lett.\ \textbf{74}, 3130 (1995); 
A.~Abada and M.~C.~Birse, Phys.\ Rev.\ D \textbf{55}, 6887 (1997); 

\bibitem{ove} 
I.~N.~Mishustin and O.~Scavenius, 
Phys.\ Rev.\ Lett.\ \textbf{83}, 3134 (1999). 

\bibitem{Scavenius:1999zc} 
O.~Scavenius and A.~Dumitru, 
Phys.\ Rev.\ Lett.\ \textbf{83}, 4697 (1999). 

\bibitem{Scavenius:2000qd} 
O.~Scavenius, A.~Mocsy, I.~N.~Mishustin and
D.~H.~Rischke, 
Phys.\ Rev.\ C \textbf{64}, 045202 (2001). 

\bibitem{Scavenius:2001bb} 
O.~Scavenius, A.~Dumitru, E.~S.~Fraga,
J.~T.~Lenaghan and A.~D.~Jackson, 
Phys.\ Rev.\ D \textbf{63}, 116003 (2001). 

\bibitem{Ignatius:1993qn}
J.~Ignatius, K.~Kajantie, H.~Kurki-Suonio and M.~Laine,
Phys.\ Rev.\ D {\bf 49}, 3854 (1994).

\bibitem{paech} 
K.~Paech, H.~Stoecker and A.~Dumitru,
Phys.\ Rev.\ C {\bf 68}, 044907 (2003); 
K.~Paech and A.~Dumitru,
Phys.\ Lett.\ B {\bf 623}, 200 (2005).

\bibitem{Aguiar:2003pp}
C.~E.~Aguiar, E.~S.~Fraga and T.~Kodama,
J.\ Phys.\ G {\bf 32}, 179 (2006).

\bibitem{polyakov} 
R.~D.~Pisarski, 
Phys.\ Rev.\ D \textbf{62}, 111501 (2000); 
A.~Dumitru and R.~D.~Pisarski,
Phys.\ Lett.\ B {\bf 504}, 282 (2001); 
A.~Dumitru and R.~D.~Pisarski, 
Nucl.\ Phys.\ A \textbf{698}, 444 (2002). 

\bibitem{explosive} 
O.~Scavenius, A.~Dumitru and A.~D.~Jackson, 
Phys.\ Rev.\ Lett.\ \textbf{87}, 182302 (2001). 

\bibitem{Csernai:1992tj}
L.~P.~Csernai and J.~I.~Kapusta,
Phys.\ Rev.\ D {\bf 46}, 1379 (1992).

\bibitem{gellmann} 
M.~Gell-Mann and M.~Levy, Nuovo Cim.\ \textbf{16}, 705
(1960); 
R.~D.~Pisarski, 
Phys.\ Rev.\ Lett.\ \textbf{76}, 3084 (1996). 

\bibitem{Fraga:1994xd}
E.~S.~Fraga and C.~A.~A.~de Carvalho,
Phys.\ Rev.\ B {\bf 52}, 7448 (1995).

\bibitem{kapusta-book} 
J.\ Kapusta, 
\textit{Finite Temperature Field Theory}
(Cambridge University Press, Cambridge, 1989).

\bibitem{licinio}
A.~Dumitru, L.~Portugal and D.~Zschiesche,
nucl-th/0502051; 
Phys.\ Rev.\ C {\bf 73}, 024902 (2006).

\bibitem{fraser}
C.~M.~Fraser,
Z.\ Phys.\ C {\bf 28}, 101 (1985); 
I.~J.~R.~Aitchison and C.~M.~Fraser,
Phys.\ Rev.\ D {\bf 31}, 2605 (1985).

\bibitem{sakita} 
Z. Su and B. Sakita, 
Phys.\ Rev.\ B \textbf{38}, 7421 (1988); 
P. K. Panigrahi, R. Ray and B. Sakita, 
{\it ibid.} \textbf{42}, 4036 (1990); 
C. A. A. de Carvalho, D. G. Barci and L. Moriconi, 
Phys.\ Rev.\ B \textbf{50}, 4648 (1994); 
D.~G.~Barci, E.~S.~Fraga and C.~A.~A.~de Carvalho,
Phys.\ Rev.\ D {\bf 55}, 4947 (1997).

\bibitem{Pisarski:1984ms} 
R.~D.~Pisarski and F.~Wilczek, Phys.\ Rev.\ D 
\textbf{29}, 338 (1984). 

\bibitem{rajaraman} 
R. Rajaraman, 
{\it Solitons and Instantons} (North-Holland, 1989).

\bibitem{yulu} 
Yu-Lu (Ed.), 
{\it Solitons and Polarons in Conducting Polymers}
(World Scientific, 1988).

\bibitem{polarons} 
D. Boyanovsky, C. A. A. de Carvalho and E. S. Fraga, 
Phys. Rev. B \textbf{50}, 2889 (1994).

\bibitem{itzykson} 
C. Itzykson and J.-B. Zuber, 
{\it Quantum Field Theory} (Dover Publications, 2006).

\bibitem{Fraga:2004hp}
E.~S.~Fraga and G.~Krein,
Phys.\ Lett.\ B {\bf 614}, 181 (2005); 
E.~S.~Fraga,
hep-ph/0510344.

\bibitem{reviews}
J. D. Gunton, M. San Miguel and P. S. Sahni, in 
\textit{Phase Transitions and Critical Phenomena} 
(Eds.: C. Domb and J. L. Lebowitz, Academic Press, London, 1983), v.~8.

\bibitem{bosonic}
T.~D.~Lee and M.~Margulies,
Phys.\ Rev.\ D {\bf 11}, 1591 (1975)
[Erratum-ibid.\ D {\bf 12}, 4008 (1975)];
A.~Bochkarev and J.~I.~Kapusta,
Phys.\ Rev.\ D {\bf 54}, 4066 (1996);
G.~W.~Carter, P.~J.~Ellis and S.~Rudaz,
III: Mesons
Nucl.\ Phys.\ A {\bf 618}, 317 (1997);
H.~C.~G.~Caldas, A.~L.~Mota and M.~C.~Nemes,
Phys.\ Rev.\ D {\bf 63}, 056011 (2001); 
A.~Mocsy, I.~N.~Mishustin and P.~J.~Ellis,
Phys.\ Rev.\ C {\bf 70}, 015204 (2004).


\end{thebibliography}
\end{document}